\documentclass{comnet}

\usepackage{multirow,booktabs,amsmath,soul,color,pdfcomment,xcolor,algorithm,algorithmic}
\usepackage[pdftex]{graphicx}
\usepackage{xcolor}
\usepackage{soul}

\graphicspath{{Graphics/}}

\begin{document}

\title{SINDyG: Sparse Identification of Nonlinear Dynamical Systems from Graph-Structured Data, with Applications to Stuart-Landau Oscillator Networks}

\shorttitle{SINDyG: Sparse Identification from Graph Data} 
\shortauthorlist{M.A. Basiri \& S. Khanmohammadi} 

\author{
\name{Mohammad Amin Basiri}
\address{Data Science and Analytics Institute, University of Oklahoma, Norman, OK, USA.}
\and
\name{Sina~Khanmohammadi$^*$}
\address{Data Science and Analytics Institute, University of Oklahoma, Norman, OK, USA.\\ School of Computer Science, University of Oklahoma, Norman, OK, USA.\email{$^*$Corresponding Author: sinakhan@ou.edu}}}

\maketitle

\begin{abstract}
{The combination of machine learning (ML) and sparsity-promoting techniques is enabling direct extraction of governing equations from data, revolutionizing computational modeling in diverse fields of science and engineering. The discovered dynamical models could be used to address challenges in climate science, neuroscience, ecology, finance, epidemiology, and beyond. However, most existing sparse identification methods for discovering dynamical systems treat the whole system as one without considering the interactions between subsystems. As a result, such models are not able to capture small changes in the emergent system behavior. To address this issue, we developed a new method called Sparse Identification of Nonlinear Dynamical Systems from Graph-structured data (SINDyG), which incorporates the network structure into sparse regression to identify model parameters that explain the underlying network dynamics. We tested our proposed method using several case studies of neuronal dynamics, where we modeled the macroscopic oscillation of a population of neurons using the extended Stuart-Landau (SL) equation and utilize the SINDyG method to identify the underlying nonlinear dynamics. Our extensive computational experiments validate the improved accuracy and simplicity of discovered network dynamics when compared to the original SINDy approach. The proposed graph-informed penalty can be easily integrated with other symbolic regression algorithms, enhancing model interpretability and performance by incorporating network structure into the regression process.}
{Sparse Regression, Model Extraction, Network Dynamics, Neural Dynamics, Stuart-Landau Model}
\\
\end{abstract}

\section{Introduction}\label{sec:introduction}
Recent breakthroughs in machine learning and data science~\cite{I2} have introduced a new phase of analyzing complex data, making it possible to identify patterns in various large datasets that are too complex for humans to understand. Yet, the creation of dynamic models that can capture the full range of behaviors within a system, especially those behaviors not observed in the initial data collection, remains a challenge ~\cite{I3}. Some of the early work in this domain includes methods introduced by researchers like Bongard and Lipson~\cite{I3}, and Schmidt and Lipson~\cite{I4}. These methods use techniques like symbolic regression and genetic programming~\cite{I5} to discover the underlying differential equations from data. 

However, these methods were held back by high computational demands and scalability issues, highlighting the need for ongoing improvements~\cite{I7}~\cite{I8}. In this domain, techniques focusing on sparse identification~\cite{paper1}, which aim to simplify the dynamics to a basic set of functions, have shown promise. For example, Sparse Identification of Nonlinear Dynamics (SINDy) ~\cite{paper1} uses sparse regression to incorporate known physical laws and partial knowledge of the system, improving the accuracy of models derived from noisy and incomplete data~\cite{I44}~\cite{SR3}. 

While SINDy as a general framework could be applied to a wide range of systems including networks ~\cite{nijholt2022emergent, topal2023reconstructing, delabays2024hypergraph}, it typically does not incorporate prior knowledge of the network structure during model identification. Instead, it treats all candidate terms equally, regardless of whether they reflect actual node-to-node connections or not. As a result, when applied to systems with many interacting components such as biological neural networks in the brain~\cite{brain}, power grids~\cite{powergrid}, social networks~\cite{I97}, and traffic systems~\cite{traffic}, SINDy may lead to the selection of spurious terms that do not represent known connectivity patterns. This could potentially yield to biased and less arcuate models that do not reflect the true dynamics of the system.


In this paper, we propose an extension of the SINDy method, termed SINDyG (Sparse Identification of Nonlinear Dynamics for Graph-structured data), which explicitly accounts for the network structure during sparse regression. By incorporating the connections between nodes, our approach uncovers not only the individual dynamics of each node but also the mechanisms by which they interact, leading to more accurate and interpretable models. 
SINDyG introduces a graph-informed regularization strategy that uses prior knowledge of the network’s adjacency structure to guide model discovery. This enables more accurate and parsimonious identification of governing equations, particularly in systems where node-to-node interactions are sparse and structured. We tested our method using several case studies of neuronal dynamics built around the extended Stuart–Landau oscillator networks, where the goal is to identify nonlinear interactions between populations of neurons. Our results indicate improvements to model accuracy and simplicity when compared to the original SINDy method. The proposed graph-informed regularization approach is a modular framework that can be easily adapted and integrated with other symbolic regression solvers, promoting broader applicability and enhanced performance.

The rest of the work is organized as follows: In Sec. II, we discuss the necessary background information such as Sparse Identification of Nonlinear Dynamics (SINDy)~\cite{paper1} and graph-structured data. In Sec. III, we explain our proposed method for data-driven modeling of complex and interconnected dynamical systems. Next, Section IV presents the results of our method using several case studies based on neuronal dynamics. Finally, we conclude the paper in Section V. 

\section{Preliminaries}
\label{Sec:Preliminaries}

The following sections provide a description of Graph-structured time series data and Sparse Identification of Nonlinear Dynamics (SINDy), which are going to be frequently used in this paper.

\subsection{Graph Structured Time Series Data}

Graphs provide a natural framework for representing systems where nodes (representing entities or states) and edges (representing interactions or connections) capture the underlying structure and dynamics of the network. Graph-structured data refers to multivariate time series where each variable corresponds to a node in a graph, and the evolution of a node depends not only on its own state but also on the states of its connected neighbors. The dynamics within a graph-structured system are governed by how changes in one node influence other nodes through their connections. This influence can be due to various factors such as physical interactions, information flow, or dependency relations. For instance, in a network of oscillators, the state of each oscillator at any given time is influenced by the states of the oscillators it is connected to, forming a complex web of interactions that evolve over time.

We represent the connections in the graph using an adjacency matrix $\mathbf{A}$, where each element $\mathbf{A}_{mn}$ denotes the presence of a connection from node $m$ to node $n$. In a directed graph, $\mathbf{A}_{mn}$ can differ from $\mathbf{A}_{nm}$, indicating that the influence from node $m$ to node $n$ is not necessarily reciprocal. Connections in the graph, represented by the edges, signify the pathways through which information or influence is transmitted between nodes. These connections can be weighted to represent the varying strengths of interactions. For example, in a neural network, a stronger synaptic connection might result in a more significant influence of one neuron's activity on another.

In contrast to generic dynamical systems, where all candidate interactions are considered equally likely, graph-structured data imposes a prior structure on variable dependencies. In standard SINDy, the sparse regression step does not account for known network topology, often leading to the inclusion of spurious interactions. In this work, we incorporate graph structure directly into the regression process via a connectivity-aware penalty matrix.

\subsection{Sparse Identification of Nonlinear Dynamics (SINDy)}

The SINDy method relies on parsimonious governing equations, which are mathematical equations that describe a system's behavior in the simplest and most concise way possible while still capturing its essential features. The method begins with the general dynamical systems form:

\begin{equation}
	\label{eq:maindynamics}
	\frac{\mathrm{d} \mathbf{x}(t)}{\mathrm{d} t}=\mathbf{F}\left(\mathbf{x}(t)\right).
\end{equation}

The function $\mathbf{F}\left(\mathbf{x}(t)\right)$ represents the dynamic constraints that specify the equations of motion of the system, and the vector $\mathbf{x}(t)$ indicates the state of a system at time $t$.

In order to compute the function $\mathbf{F}$ from data, a time series of system's state vector $\mathbf{x}(t)$, which contains all the state variables, is collected. The derivatives $\mathbf{\dot{x}}(t)$ are then measured or numerically approximated and arranged in the following matrix:

\begin{equation}
	\label{eq:xdot}
	\resizebox{.5\hsize}{!}{$
		\dot{\mathbf{X}}=\left[\begin{array}{c}
			\dot{\mathbf{x}}^T\left(t_1\right) \\
			\dot{\mathbf{x}}^T\left(t_2\right) \\
			\vdots \\
			\dot{\mathbf{x}}^T\left(t_T\right)
		\end{array}\right]=\left[\begin{array}{cccc}
			\dot{x}_1\left(t_1\right) & \dot{x}_2\left(t_1\right) & \cdots & \dot{x}_K\left(t_1\right) \\
			\dot{x}_1\left(t_2\right) & \dot{x}_2\left(t_2\right) & \cdots & \dot{x}_K\left(t_2\right) \\
			\vdots & \vdots & \ddots & \vdots \\
			\dot{x}_1\left(t_T\right) & \dot{x}_2\left(t_T\right) & \cdots & \dot{x}_K\left(t_T\right)
		\end{array}\right]$}.
\end{equation}

Afterwards, a library $\mathbf{\Theta}(\mathbf{X})$ consisting of candidate nonlinear functions of the columns of $\mathbf{X}$ is constructed.  For instance, $\mathbf{\Theta}(\mathbf{X})$ may consist of constant, polynomial, and trigonometric terms:

\begin{equation}
	\label{eq:library}
	\resizebox{.5\hsize}{!}{$
		\mathbf{\Theta}(\mathbf{X})=\left[\begin{array}{cccccccc}
			\mid & \mid & \mid & \mid & & \mid & \mid\\
			1 & \mathbf{X} & \mathbf{X}^{P_2} & \mathbf{X}^{P_3} & \cdots & \sin (\mathbf{X}) & \cos (\mathbf{X}) & \cdots \\
			\mid & \mid & \mid & \mid & & \mid & \mid &
		\end{array}\right]$},
\end{equation}

where, higher polynomials are shown as $\mathbf{X}^{P_2}$, $\mathbf{X}^{P_3}$, $\cdots$. For example, $\mathbf{X}^{P_2}$ denotes the quadratic nonlinearities in the state variables $\mathbf{x}$. A potential function for the right-hand side of Eq~\ref{eq:maindynamics} is represented by each column of $\mathbf{\Theta}(\mathbf{X})$ in Eq~\ref{eq:library}. The values of each column are often normalized to prevent numerical instability and ensure that features contribute equally to the model fitting process. Since only a small number of these terms are active in each row of $\mathbf{F}$, a sparse regression problem will be established to find the sparse vectors of coefficients $\mathbf{\Xi}=\left[\begin{array}{llll}\boldsymbol{\xi}_1 & \boldsymbol{\xi}_2 & \cdots & \boldsymbol{\xi}_K\end{array}\right]$ that determine which nonlinear terms are active:

\begin{equation}
	\label{eq:sparsereg}
	\dot{\mathbf{X}} = \mathbf{\Theta}(\mathbf{X})\mathbf{\Xi}.
\end{equation}

For each column $\boldsymbol{\xi}_k$ of $\mathbf{\Xi}$, there is a sparse vector of coefficients that determine which terms are active in the righthand side for one of the row equations $\dot{\mathbf{x}}_k=\mathbf{F}_k\left(\mathbf{x}\right)$ in Eq~\ref{eq:maindynamics}. After $\mathbf{\Xi}$ is identified, a model of each row of the governing equations can be built as follows:

\begin{equation}
	\label{eq:sparsereg2}
	\dot{\mathbf{x}}_k=\mathbf{F}_k\left(\mathbf{x}\right) = \mathbf{\Theta}({\mathbf{x}}^T)\boldsymbol{\xi}_k.
\end{equation}

Unlike $\mathbf{\Theta}(\mathbf{X})$, which is a data matrix, $\mathbf{\Theta}({\mathbf{x}}^T)$ is a vector of symbolic functions of elements of ${\mathbf{x}}$. As a result,

\begin{equation}
	\label{eq:sparsereg3}
	\dot{\mathbf{x}}=\mathbf{F}\left(\mathbf{x}\right) = \mathbf{\Xi}^T\left(\mathbf{\Theta}({\mathbf{x}}^T)\right)^T.
\end{equation}

To obtain the sparse vector of coefficients $\boldsymbol{\xi}_k$ for the kth row equation, each column of $\dot{\mathbf{X}}$ in Eq~\ref{eq:sparsereg} requires a distinct optimization. Furthermore, the optimization is given by

\begin{equation}
	\label{eq:sindyOpt}
	\min _{\mathbf{\Xi}} \frac{1}{2}\|\dot{\mathbf{X}}-\mathbf{\Theta}(\mathbf{X}) \mathbf{\Xi}\|^2+\lambda \mathbf{R}(\mathbf{\Xi}),
\end{equation}

where $\lambda$ is a hyper-parameter that controls the regularization's strength and $\mathbf{R}(.)$ is a regularizer that imposes sparsity or other constraints on the solution. There are several well-known techniques for Eq~\ref{eq:sindyOpt} when $\mathbf{R}(.)$ is convex. The typical technique is to use $\mathbf{R}(.)$ to represent the sparsity-promoting $L1$ norm, which is a convex relaxation of the $L0$ norm. While LASSO~\cite{Lasso} is commonly used to solve SINDy due to its convex formulation, it has several limitations such as bias in coefficient estimation. Hence, the SINDy algorithm employs sequential thresholded least squares (STLSQ), which given a parameter $\eta$ that specifies the minimum magnitude for a coefficient in $\mathbf{\Xi}$, performs a least squares fit and then zeros out all the coefficients with magnitude below the threshold. 
This process of fitting and thresholding is repeated until convergence.
The convergence properties of the STLSQ algorithm have been discussed in \cite{zhang2019convergence}, which shows that, under assumptions such as exact sparsity of the true model and a well-conditioned feature matrix, the procedure converges to a stable sparse solution.
The SINDy algorithm has recently been integrated into a python package PySINDy~\cite{Pysindy}.

\section{Materials and Method}

\begin{figure}[h]
	\centering
	\includegraphics[width=\linewidth]{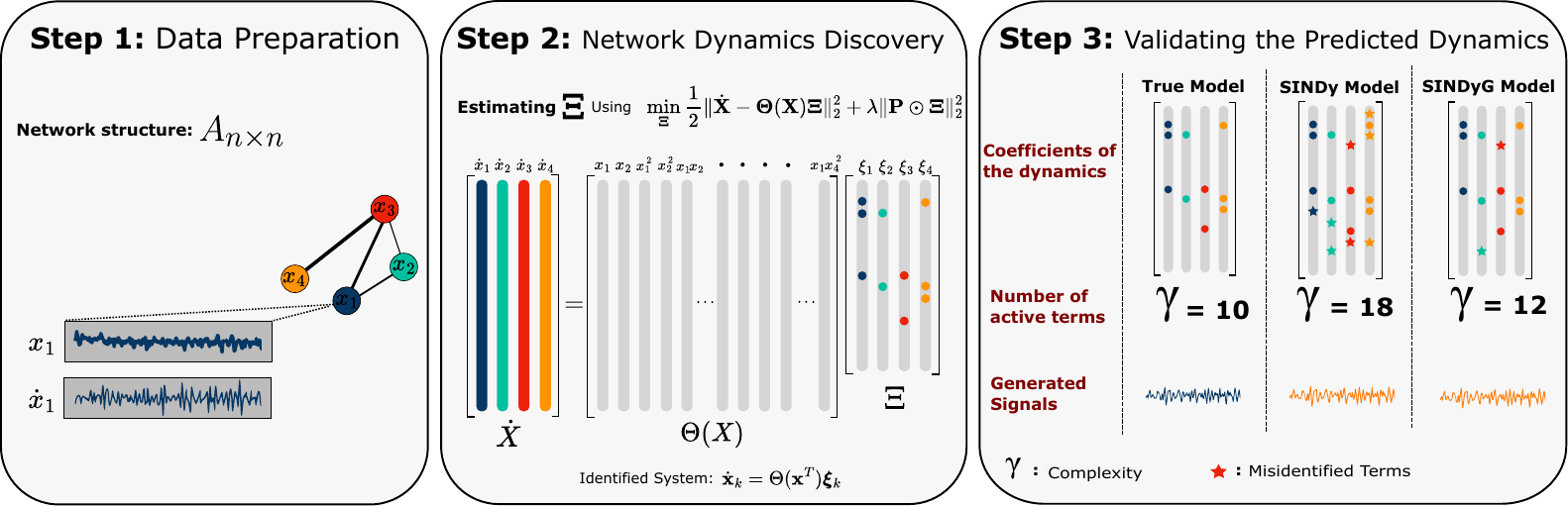}
	\caption{Overview of the proposed framework for network dynamics discovery: (Step 1) Data preparation involves collecting time-series data $\mathbf{X}$ from a graph-structured system. (Step 2) The SINDyG algorithm constructs a library of basis functions $\mathbf{\Theta}(\mathbf{X})$, applying sparse regression via the objective function and STLSQ procedure to identify the optimal sparse coefficient matrix $\mathbf{\Xi}$ that satisfies $\dot{\mathbf{X}}=\mathbf{\Theta}(\mathbf{X})\mathbf{\Xi}$ while incorporating the network structure. (Step 3) The predicted dynamics are validated using appropriate metrics to evaluate model performance ~\cite{paper1}~\cite{sindypic}.
    \newline\newline
    \textbf{Alt Text:} A three-panel flowchart illustrating the SINDyG framework. Step 1 shows data preparation with a network graph and time-series plots. Step 2 depicts network dynamics discovery with a matrix equation. Step 3 validates the model by comparing coefficients, number of terms and generated signals of the True, SINDy, and SINDyG models.
}
	\label{fig:framework}
\end{figure}

The SINDyG method is based on the SINDy algorithm that was introduced in the previous section. In SINDyG, shown in figure~\ref{fig:framework}, the time histories of the state variables $\mathbf{X}$ are collected, their derivatives are calculated, and the structure of the network is identified. Next, a library of functions of the state variables, $\mathbf{\Theta}(\mathbf{X})$, is constructed. This feature library is used to find the fewest terms needed to satisfy $\dot{\mathbf{X}}=\mathbf{\Theta}(\mathbf{X}) \mathbf{\Xi}$ while taking into account the network structure. The few active entries in the vectors of $\mathbf{\Xi}$ are identified using sparse regression~\cite{paper1}. Here, we provide details of each step of the SINDyG method.\\ 

\subsection{Data Preparation}

Data preparation is a crucial step in the discovery of dynamical systems using the SINDyG method. This preparation involves gathering and organizing three key components: an adjacency matrix, time series data, and their derivatives. These components collectively provide the necessary information to uncover the underlying equations governing the system's behavior.\\

\subsection{Network Dynamics Discovery}

In this step, we aim to uncover the underlying dynamical equations governing each node in the network. We begin by constructing a library matrix, $\mathbf{\Theta}(\mathbf{{X}})$, using a predefined set of basis functions. These basis functions represent potential terms that could contribute to the dynamics~\cite{paper1}. Next, we employ a sequentially thresholded least squares algorithm (STLSQ)~\cite{paper1} to determine the coefficients $\mathbf{\Xi}$ corresponding to these basis functions. This is achieved by minimizing the following objective function:

\begin{equation}
	\label{eq:sindyGOpt}
	\min _\mathbf{\Xi} \frac{1}{2}\|\dot{\mathbf{X}}-\mathbf{\Theta}(\mathbf{X}) \mathbf{\Xi}\|_2^2+\lambda\|\mathbf{P}\odot\mathbf{\Xi}\|_2^2.
\end{equation}

The parameter $\lambda$ controls the amount of regularization. The term $\mathbf{P}\odot\mathbf{\Xi}$ represents the element-wise product of $\mathbf{P}$ and $\mathbf{\Xi}$, while the penalty matrix $\mathbf{P}$ leverages the network's adjacency matrix $\mathbf{A}$ to introduce a graph-aware penalty. Unlike the original SINDy method, which applies uniform penalties to all terms in the candidate function library, SINDyG introduces a penalty matrix 
$\mathbf{P}$ that biases the regression toward selecting terms consistent with the known graph structure. The intuition behind the penalty matrix $\mathbf{P}$ is to determine the relative importance of different terms in the library based on the network's connectivity. By encouraging coefficients linked to strongly connected nodes and discouraging those linked to weakly connected nodes, this penalty term guides the algorithm toward solutions that ensure the discovered dynamics are consistent with the underlying network topology.



In order to calculate this penalty, we iterate through each candidate term, $c$, in the library of functions. For each term, we identify the source indices,  $S_c$, representing the state variables present in term $c$. We then determine the sink indices, $D_c$, which are the state variables reachable from $S_c$. For each sink index $k \in D_c$, we compute the mean connectivity, $m_{ck}$, between the source indices and sink index $k$ using the adjacency matrix $\mathbf{A}$.
For a candidate term 
$c$ involving source variables  $S_c$, we define the mean connectivity to sink node $k$ as:
\begin{equation}
	\label{eq:mck}
	m_{ck}=\frac{1}{\left|S_c\right|} \sum_{i \in S_c} \mathbf{A}[i, k].
\end{equation}

\begin{figure}[h]
	\centering
	\includegraphics[width=0.72\linewidth]{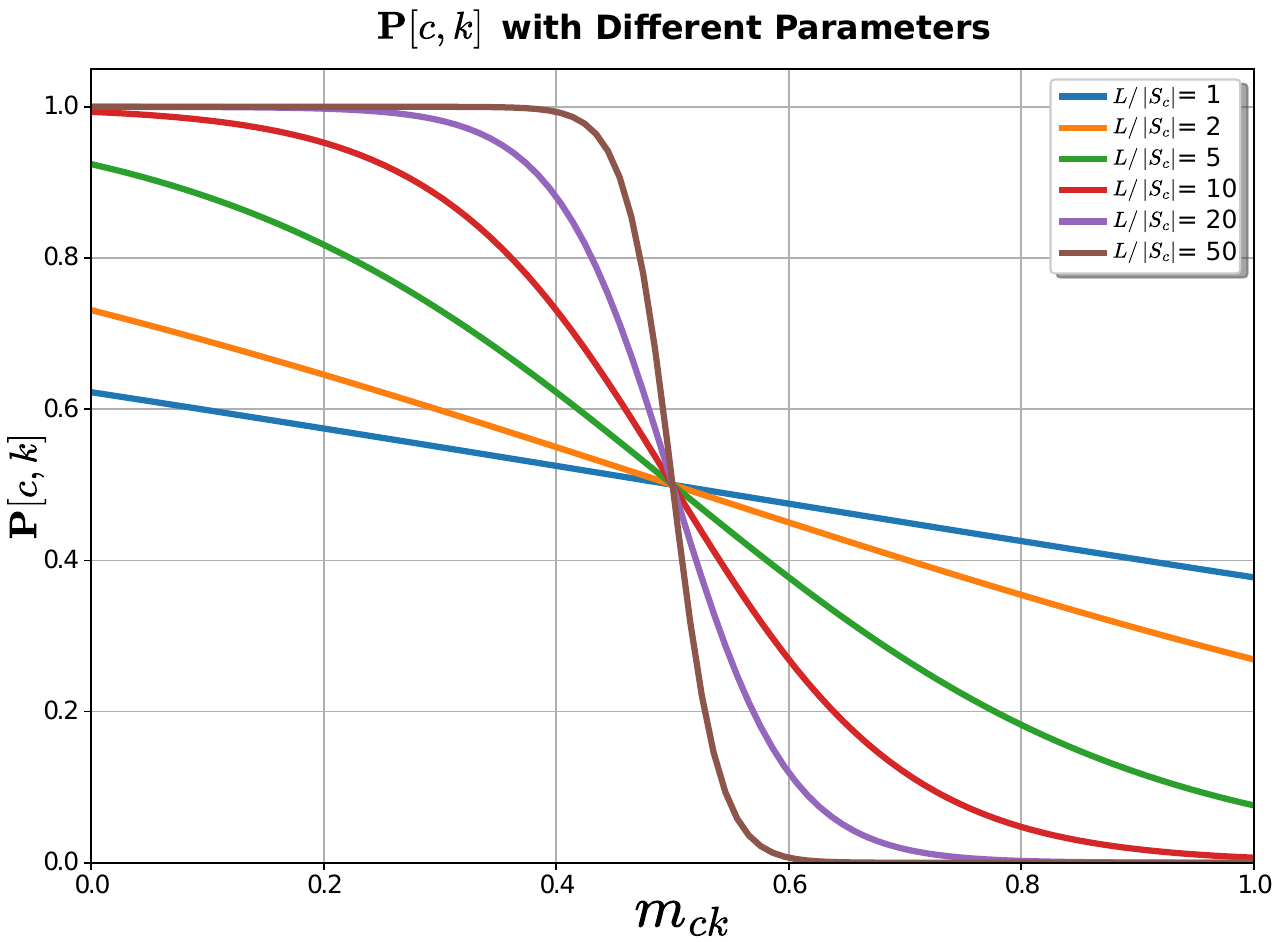}
	\caption{Comparison of $\mathbf{P}[c,k]$ with different values of $L/\left|S_c\right|$ and $m_{c k}$
    \newline\newline
    \textbf{Alt Text:}  A line graph showing the relationship between the penalty value P[c,k] and mean connectivity $m_{c k}$. Multiple curves illustrate how the parameter $L/\left|S_c\right|$ affects the shape of the penalty function, transitioning from a gentle slope to a steep, step-like function as the parameter value increases.}
	\label{fig:sigmoid_comparison}
\end{figure}

This captures the average influence of the source variables on node 
$k$ under the network topology.
The penalty matrix is then calculated using a formula similar to the sigmoid formula:

\begin{equation}
	\label{eq:f}
	\mathbf{P}[c, k]=\frac{1}{1+\exp\left(\left(L/\left|S_c\right|\right)\cdot \left(m_{c k}-0.5\right) \right)},
\end{equation}

where $L$ is a parameter for adjusting the shape of the output of this formula. As we increase $L$, the behavior of the equation becomes more similar to a step function. Conversely, if we reduce $L$, the equation's behavior resembles a ramp function. This formulation ensures that terms with low mean connectivity between source and sink variables receive a higher penalty (closer to 1), while the penalty is scaled by the number of source variables, $\left|S_c\right|$, to account for terms involving multiple state variables. The shape of the value of $\mathbf{P}[c,k]$ based on the value of $L/\left|S_c\right|$ is illustrated in figure~\ref{fig:sigmoid_comparison}.

To solve the optimization problem shown in Eq.~\ref{eq:sindyGOpt} for each state variable $k$, we perform a series of transformations that allow us to leverage the standard ridge regression framework. First, we scale the feature matrix $\mathbf{\Theta}(\mathbf{X})$ and the coefficient vector $\boldsymbol{\xi}_k$ to simplify the regularization term. Specifically, we define the transformed variables $\mathbf{\Theta}^{\prime}(\mathbf{X})$ and ${\boldsymbol{\xi}_k^{\prime}}$ as follows: 
\[
\mathbf{\Theta}^{\prime}(\mathbf{X})
=
\mathbf{\Theta}(\mathbf{X}) \cdot \operatorname{diag}(1 / \boldsymbol{p}_k),
\quad
{\boldsymbol{\xi}_k^{\prime}}
=
\operatorname{diag}(\boldsymbol{p}_k) \cdot \boldsymbol{\xi}_k,
\]
where each column $c$ of $\mathbf{\Theta}(\mathbf{X})$ is divided by the corresponding element $\boldsymbol{P}_{ck}$, and ${\boldsymbol{\xi}_k^{\prime}}=\operatorname{diag}(\boldsymbol{p}_k) \cdot \boldsymbol{\xi}_k$, which scales each coefficient in $\boldsymbol{\xi}_k$ by the corresponding element in $\boldsymbol{p}_k$ which is column $k$ of $\mathbf{P}$.
These substitutions transform the original problem into the equivalent form $\|\dot{\mathbf{x}}_k-\mathbf{\Theta}^{\prime}(\mathbf{X}) {\boldsymbol{\xi}_k^{\prime}}\|_2^2+\lambda\|{\boldsymbol{\xi}_k^{\prime}}\|_2^2$.

In this transformed space, the problem becomes a standard ridge regression, which can be solved efficiently using existing algorithms. After obtaining the solution ${\boldsymbol{\xi}_k^{\prime}}$ using the sequentially thresholded least squares algorithm, we revert to the original coefficient vector $\boldsymbol{\xi}_k$ by dividing ${\boldsymbol{\xi}_k^{\prime}}$ by $\boldsymbol{p}_k$, ensuring that the final coefficients correspond to the original problem's scaling. This approach effectively incorporates the custom regularization term $\lambda\|\boldsymbol{p}_k\odot\boldsymbol{\xi}_k\|_2^2$ into the ridge regression framework without altering the underlying algorithm. By scaling the feature matrix and coefficients appropriately, we maintain the efficiency of the standard ridge regression solver while addressing the modified regularization objective.

While Eq.~\ref{eq:sindyGOpt} presents a ridge regression objective incorporating a graph-informed penalty matrix 
$\mathbf{P}$, sparsity is not enforced directly via 
$L0$ norm. Instead, we employ a modified version of the Sequentially Thresholded Least Squares (STLSQ) algorithm, which iteratively alternates between solving Eq.~\ref{eq:sindyGOpt} and applying a hard threshold to small coefficients in  $\mathbf{\Xi}$. This results in a sparse coefficient matrix over iterations. This iterative procedure ensures that the discovered model aligns with the known network structure while maintaining interpretability and computational efficiency. The complete graph-aware STLSQ procedure is summarized below.

\begin{algorithm}[H]
\caption{Summary of Graph-aware STLSQ Procedure in SINDyG}
\begin{algorithmic}[1]
\REQUIRE Time series data $X$, derivatives $\dot{X}$, adjacency matrix $A$, threshold $\eta$, penalty parameter $\lambda$
\ENSURE Sparse coefficient matrix $\Xi$
\STATE Construct the library matrix $\Theta(X)$
\STATE Compute penalty matrix $P$ using mean connectivity based on $A$
\STATE Initialize coefficient matrix $\Xi$
\REPEAT
    \FOR{each column $k$ of $\Xi$}
        \STATE Scale $\Theta(X)$ and $\xi_k$ using penalty vector $p_k$ (column $k$ of $P$)
        \STATE Solve ridge regression subproblem:
        \[
        \min_{\xi_k'} \|\dot{x}_k - \Theta'(X) \xi_k'\|_2^2 + \lambda \|\xi_k'\|_2^2
        \]
        \STATE Revert scaling: $\xi_k \leftarrow \xi_k' / p_k$
        \STATE Threshold small entries in $\xi_k$: set elements $< \eta$ to zero
    \ENDFOR
\UNTIL{convergence of $\Xi$}
\RETURN $\Xi$
\end{algorithmic}
\end{algorithm}

\subsection{Validating the Predicted Dynamics}

Once the underlying equations governing a system have been identified using techniques such as SINDy or SINDyG, a comprehensive validation process is essential to assess the model's performance. This involves evaluating the predicted model from multiple perspectives. Key metrics employed in this study include the model complexity index, the coefficient Mean Absolute Error (MAE), the generated signal accuracy, and training time. 

The model complexity index ($\gamma$) assesses the simplicity of the identified equations based on the number of active terms. A lower $\gamma$ value signifies a more parsimonious model, potentially easier to interpret and generalize. If the true mathematical equation describing the dynamics is known, the model complexity index could also be used to compare the complexity of the identified and true models, allowing us to navigate the trade-off between accuracy and interpretability for more informative and reliable models. By considering model complexity with other metrics, we can ensure that the predicted model captures the essence of the system dynamics while remaining interpretable.

The coefficient mean absolute error (MAE) evaluates how closely the learned coefficients match the true coefficients. It is computed as the average absolute difference between the predicted coefficients and the true coefficients. We chose MAE over other metrics such as L-infinity norm to emphasize interpretability and robustness. While the L-infinity norm is useful for identifying the single worst-case deviation, it can be dominated by a single outlier and does not reflect the model's global accuracy across all coefficients. In contrast, MAE provides a clear indication of the average deviation per term and is less influenced by outliers, which is desirable in sparse models where most terms are zero and a few incorrect activations can otherwise dominate the score. Mathematically, the coefficient MAE is defined as:

\begin{equation}
	\label{eq:CMAI}
	\text{Coefficient MAE} = \frac{1}{KC} \sum_{k=1}^{K}\sum_{c=1}^{C}\left|\boldsymbol{\xi}_{k,c}-\widehat{\boldsymbol{\xi}}_{k,c}\right|,
\end{equation}

where $K$ is the number of dynamical equations, $C$ represents the number of possible candidate functions for each equation, $\boldsymbol{\xi}_{k,c}$ denotes the $c^{th}$ predicted coefficient for the $k^{th}$ equation, and $\widehat{\boldsymbol{\xi}}_{k,c}$ signifies the true coefficient. A lower coefficient MAE value indicates a closer match between the predicted and true coefficients, implying higher accuracy for the predicted dynamical model. Conversely, a higher coefficient MAE value suggests a larger discrepancy between the predicted and true values, signifying a potentially less accurate coefficient prediction. This method requires the availability of true coefficients. When the underlying governing equations are not readily available, alternative references such as values obtained from established physical laws or high-fidelity simulations can be employed.

The model complexity index ($\gamma$) used throughout our experiments is defined as the number of active (nonzero) terms in the coefficient matrix $\mathbf{\Xi}$. While related to sparsity, this measure is invariant to the total number of candidate terms and provides a direct and interpretable metric of model complexity. A coefficient matrix with a few active terms corresponds to a lower $\gamma$, indicating a simpler and more interpretable model. Similarly, the coefficient MAE is also indirectly linked to the number of terms: a higher coefficient MAE often arises when additional spurious terms are incorrectly activated in $\mathbf{\Xi}$. This observation aligns with prior findings which noted that models with poor predictive performance tend to exhibit higher complexity and reduced sparsity~\cite{Kaheman}. By reporting both $\gamma$ and coefficient MAE, we capture how closely the learned dynamics match the ground truth in both structure and accuracy.

The generated signal accuracy measure focuses on comparing the trajectories of state variables, specifically their derivatives, between the observed signal and the predicted signal. Two established metrics are employed for this comparison: R-squared ($R^2$) and Mean Squared Error (MSE). The $R^2$ quantifies the proportion of variance in the observed signal's derivatives explained by the predicted signal's derivatives:

\begin{equation}
	R^2=1-\frac{\sum_{k={1}}^{K}\sum_{i={1}}^{T}\left(\widehat{\dot{\mathbf{x}}_k}\left(t_i\right)-\dot{\mathbf{x}}_k\left(t_i\right)\right)^2}{\sum_{k={1}}^{K}\sum_{i=1}^T\left(\dot{\mathbf{x}}_k\left(t_i\right)-\bar{\dot{\mathbf{x}}}_k\left(t_i\right)\right)^2}.
\end{equation}

A higher $R^2$ value indicates a stronger correlation between the two, suggesting the method effectively captures the dynamics governing the system. Conversely, a lower $R^2$ value implies a weaker correlation, potentially indicating limitations in the method's ability to accurately predict the system's behavior.

Mean Squared Error (MSE) complements $R^2$ by measuring the average squared difference between the observed and predicted signal's derivatives. A lower MSE signifies a closer match between the two, indicating higher accuracy in the predicted signal. Conversely, a higher MSE suggests a larger discrepancy, revealing potential shortcomings in replicating the observed or true system dynamics. MSE is defined as:

\begin{equation}
	\text{MSE}=\frac{1}{KT}\sum_{k={1}}^{K}\sum_{i={1}}^{T}\left(\widehat{\dot{\mathbf{x}}_k}\left(t_i\right)-\dot{\mathbf{x}}_k\left(t_i\right)\right)^2.
\end{equation}

Finally, the training time quantifies the computational efficiency of the model identification process. This is especially important for complex dynamical systems where the sparse regression models must search through a large library of candidate terms to discover an optimal sparse representation of the system dynamics.

\section{EXPERIMENTS AND ANALYSIS}

In this section, we will first introduce the different types of synthetic data used in this study and then compare the results of our proposed method with conventional SINDy approach. 

\subsection{Dataset}

We have used a neuronal dynamics model to test and validate our proposed SINDyG method. At the neural level, the synchronized activity of large number of neurons can give rise to macroscopic oscillations~\cite{oscilatory}. These oscillations are an essential aspect of brain function, reflecting coordinated neuronal activity that can be modeled in various ways~\cite{SL1}. One effective approach to model this oscillatory activity is using the Stuart-Landau (SL) equation ~\cite{SL}. The SL equation describes the behavior of a nonlinear oscillating system and could be used to model the internal dynamics of a group of neurons, capturing the essence of their collective behavior~\cite{SL}.

The SL equation in our study is defined as:
\begin{equation}
	\label{eq:SL}
	\dot{z}=\left(\sigma+\mathrm{i} \omega-|z|^2\right)z,
\end{equation}
where $z$ is a complex-valued state variable representing the activity of a neural population, $\sigma$ is the parameter that dictates the growth rate of the oscillations, and $\omega$ is the parameter that determines the oscillation frequency. For $\sigma>0$, the model exhibits sustained oscillatory behavior with frequency $\omega$. This formulation captures the fundamental dynamics of neuronal oscillations within a population. To extend the SL equation for modeling interactions between neural populations, coupling dynamics is introduced. The extended model incorporates the effects of other neural populations or external inputs, represented by:

\begin{equation}
	\label{eq:eSL}
	\dot{z}_{n}=\left(\sigma+\mathrm{i} \omega-|z_{n}|^2\right)z_{n} + kz_{n}z_{m}.
\end{equation}
In this equation, 
${z}_{n}$
denotes the state of the 
$n^{th}$ neural population, and 
${z}_{m}$
represents the dynamics of another interacting neural population. The coupling term 
$k {z}_{n} {z}_{m}$
captures the influence of the 
$m^{th}$ population on the 
$n^{th}$ neural population. This interaction term allows the model to account for the complex dynamics resulting from neural interactions. Each node's dynamics can be illustrated by a complex state variable, $z=x+iy$, which inherently represents a two-dimensional system due to its real and imaginary components ($x$ and $y$). By separating the real and imaginary components, we can decompose the complex dynamics into two independent equations, one governing the evolution of $x$ and the other for $y$. This transformation allows us to analyze the system using familiar real-valued equations, effectively eliminating the need to directly work with complex quantities. By using this extended SL model, we can simulate the neural population dynamics under various conditions. Starting from an initial condition, we can calculate how the system evolves over time. This approach enables the generation of synthetic data that reflects the true dynamics of neuronal oscillations, which could be used to test our proposed method. We have considered a simple and more general case to validate our method. \\ 

\subsubsection{Simple Oscillatory Activity}

\begin{figure}[h]
	\centering
	\includegraphics[width=0.6\linewidth]{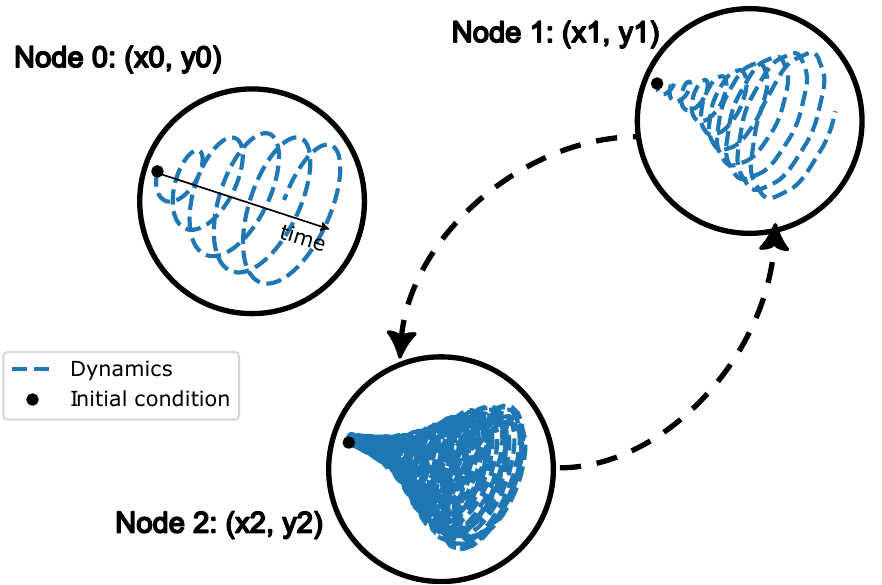}
	\caption{Example of generated neuronal dynamics data to test SINDyG. The data is based on a simple network with three nodes where nodes 1 and 2 are connected. The parameters of the SL equation for this model include: $\alpha=0.2$ for all nodes, and $\omega=\{\pi/2,\pi, 8\pi \}$, for nodes 0,1, and 2.
    \newline\newline
    \textbf{Alt Text:} A diagram of a simple network of three nodes. Node 0 is isolated, while nodes 1 and 2 are interconnected with dashed arrows. Inside each node's circle is a visual representation of its dynamic trajectory over time, starting from an initial point.}
	\label{fig:example3}
\end{figure}

In this section, the testing data is generated using the described SL oscillator framework in \ref{eq:eSL} for a simple case of graph-structured time series. In this case, as shown in figure~\ref{fig:example3}, there are three interconnected nodes, each representing the dynamics of a group of neurons. Two of the nodes are connected, and one of them is isolated from the others. These nodes generate oscillatory behavior when starting from an initial condition. This oscillatory behavior is modeled by the SL equation and can be decomposed into its real and imaginary parts. the connectivity matrix of this simple case is $\mathbf{A} = \left[\begin{array}{ccc}
	0&0&0 \\
	0&0&1 \\
	0&1&0
\end{array}\right]$.\\

\subsubsection{General Oscillatory Activity}

In this section, the dataset is generated in a flexible format to accommodate varying network sizes and structures. The number of nodes within the network is variable, and the connections between nodes adhere to the rules of either Erdős-Rényi (ER)~\cite{ER} or scale-free (SF)~\cite{SF} graph models. ER graphs feature randomly distributed edges which means each possible edge between pairs of nodes is included independently with a fixed probability. In contrast, the SF network starts from a small core of fully connected nodes, sequentially adding new nodes that are connected to existing nodes with a probability proportional to their degree, which at the end produce a power-law degree distribution with a few highly connected hubs. These network generation schemes allow us to simulate networks with varying topological properties and evaluate the robustness of the proposed SINDyG method under structurally diverse conditions.

After the random generation of the graph structure, a coupling coefficient is assigned to each available edge. This coefficient represents the strength of the connection between nodes and quantifies how much the dynamics of one node influence another. Additionally, random parameters are chosen to govern the internal dynamics of each individual node. More specifically, the internal dynamics of each node are governed by the Stuart-Landau equation, with coupling terms determined by the adjacency matrix. Initial conditions were assigned by sampling each node’s state with a completely random phase and an amplitude drawn uniformly from the interval $[0, 1]$. The time step size $dt$ was set to $0.01$ seconds. Random coupling weights between connected nodes were assigned uniformly from the interval $[0, 5]$. Frequencies for each oscillator were sampled uniformly from the range $[0.1, 2.0]$. This approach allows for the exploration of a wide range of network configurations and dynamical behaviors within the dataset.

\subsection{Results and Analysis}
\subsubsection{Simple Case Results}

\begin{figure}[h]
	\centering
	\includegraphics[width=0.75\linewidth]{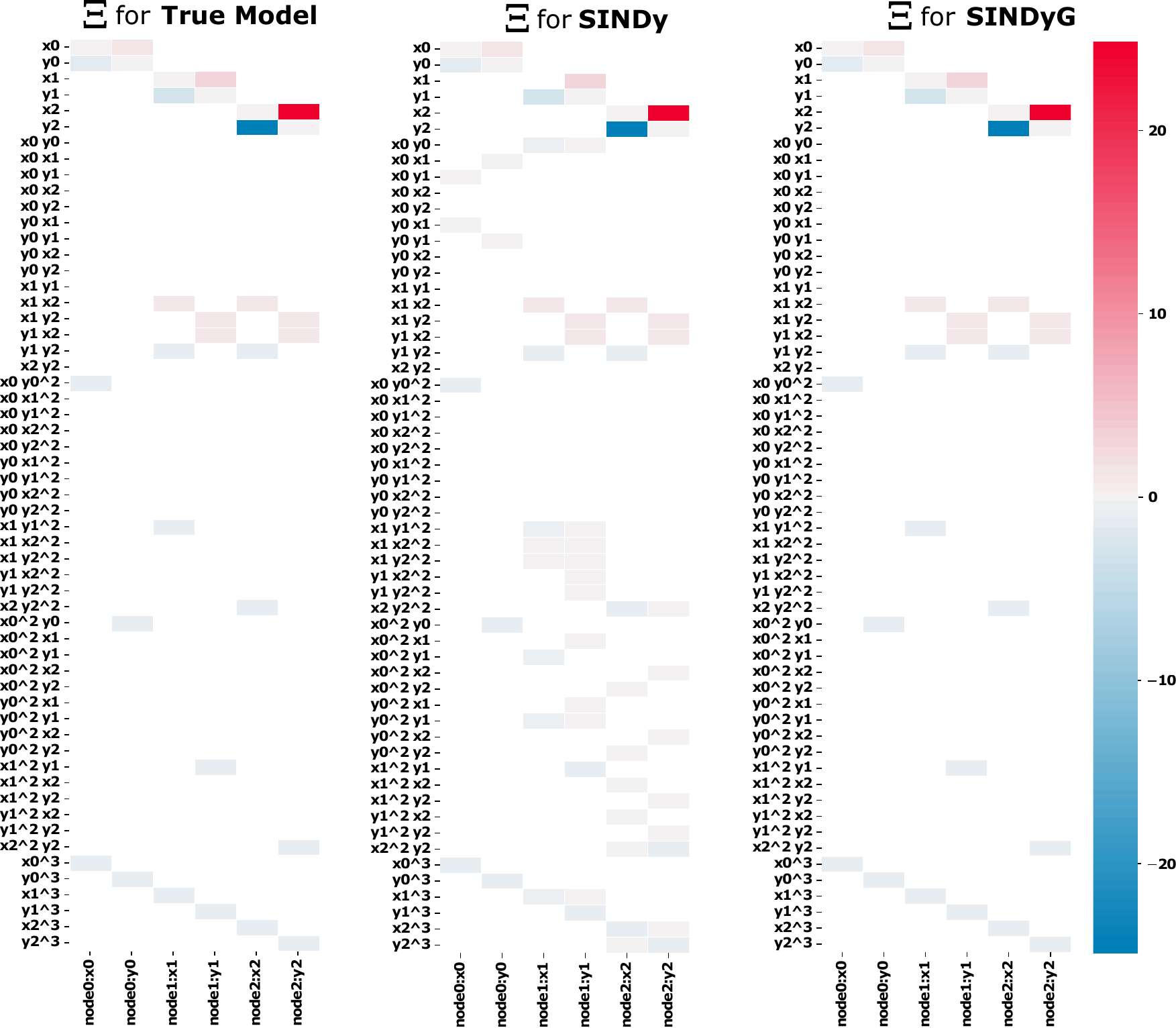}
	\caption{The heatmap showing the coefficients of extracted dynamical models using SINDy and SINDyG. The SINDyG extracted a much simpler model ($\gamma=32$), when compared to the SINDy with $\gamma=62$.
        \newline\newline
    \textbf{Alt Text:} Three parallel heatmaps comparing the coefficient matrices for the True Model, the SINDy model, and the SINDyG model. The heatmaps for the True Model and SINDyG are very similar and sparse, while the heatmap for the SINDy model shows many more active (colored) coefficients, indicating higher complexity.}
	\label{fig:heatmap}
\end{figure}

In this section, we delve into the results obtained from the simple case graph introduced earlier. This graph consists of three interconnected nodes, each governed by a specific dynamic. The true underlying model that generated the training data is known and visualized in figure~\ref{fig:heatmap}. The heatmap columns represent the coefficients for each state variable's first-order differential equation, with higher values indicating stronger contributions from specific candidate terms. For instance, the first column reveals that only four candidate terms are active in the true model for that state variable. 

In order to test our proposed method, we generated training data using the true dynamics in simple case study. This training data comprises a single trajectory, starting from an initial point and capturing the system's state evolution over a 20-second interval. This data was then used to train both SINDy and SINDyG models to uncover the underlying equations governing the system's dynamics. For this training data, the SINDyG model achieved an R-squared score of $0.99999927$ and an MSE of $0.00001297$, outperforming the SINDy model which yielded an $r^2$ score of $0.99999572$ and an MSE of $0.00001352$. Furthermore, we observed a slight advantage in computational efficiency for SINDyG. The time required for discovering the underlying equation using SINDyG was $0.04615$ seconds, compared to $0.04858$ seconds for SINDy.

Figure~\ref{fig:heatmap} presents a side-by-side comparison of the coefficients from the true model, the SINDy predicted model, and the SINDyG predicted model. The number of colored elements in each heatmap indicates the complexity of the model, with more active terms leading to higher complexity. A visual inspection reveals that the SINDy predicted model has a higher number of active terms compared to the true model, resulting in greater complexity ($\gamma=62$). In contrast, the SINDyG predicted model closely resembles the true dynamics and is more sparse compared to the SINDy predicted model and the complexity for both the true and SINDyG predicted model is $\gamma=32$.  We further quantify this similarity using the coefficient MAE, which measures the difference between the predicted and true coefficients. In this case, the SINDyG predicted model achieves a low coefficient MAE score of $0.0033$ outperforming the SINDy predicted model's coefficient MAE of $0.0155$.

\begin{figure}[h]
	\centering
	\includegraphics[width=0.75\linewidth]{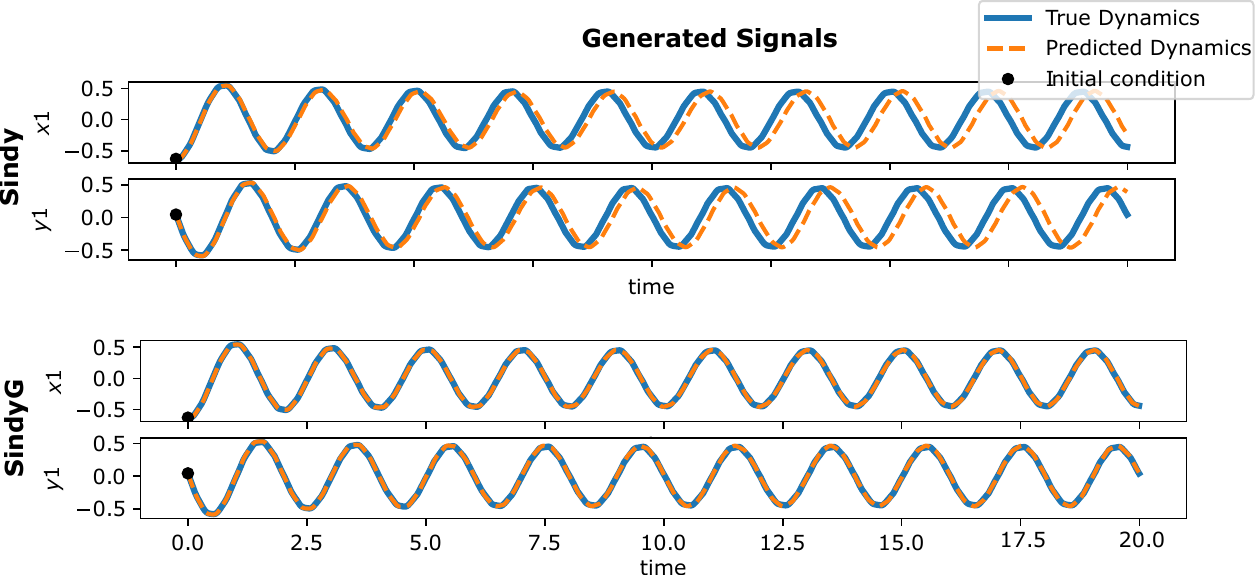}
	\caption{True and predicted trajectories of node 1 in simple case study from the same initial condition for SINDy and SINDyG. The $r^2$ score for SINDyG was 0.9999997, which is higher than the  $r^2$ score of 0.9997024 for SINDy.
    \newline\newline
    \textbf{Alt Text:} Two sets of line graphs comparing time-series signals. The top set shows the SINDy model's predicted dynamics (dashed orange line) deviating slightly from the true dynamics (solid blue line). The bottom set shows the SINDyG model's prediction almost perfectly overlapping with the true dynamics.}
	\label{fig:compare}
\end{figure}

To further validate our models, we generated new trajectories by simulating the system from random initial conditions. These trajectories were not used in the discovery process and served as unseen test data.
Figure~\ref{fig:compare} compares these generated signals with the trajectories produced using the true model. For a randomly generated new trajectory, the SINDyG model achieved $r^2$ score of $0.99999935$ and MSE of $0.00001736$, outperforming the SINDy model which yielded an $r^2$ score of 0.99988687 and MSE of 0.00009717. The R-squared score and Mean Squared Error (MSE) demonstrate that the model predicted by SINDyG outperforms the SINDy predicted model. The results are also evident in figure ~\ref{fig:compare}, where the signals generated by the SINDyG predicted model closely match those of the true model.\\

\subsubsection{General Case Results} 

\begin{figure*}[p]
	\centering
	\includegraphics[width=\textwidth]{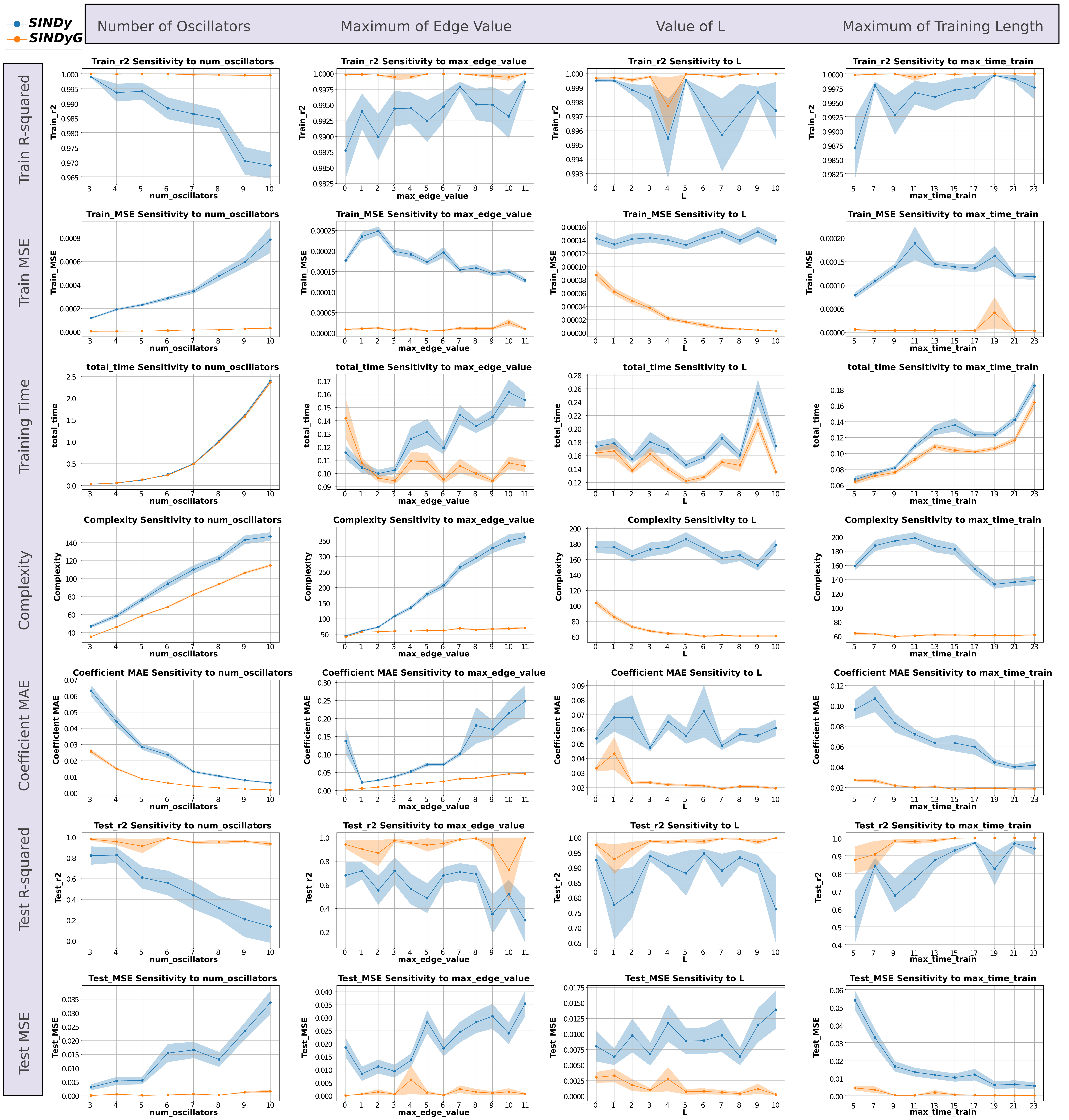}
	\caption{Sensitivity analysis of different parameters in the general case. The results are based on 100 repetitions and show the mean and standard error for different performance metrics.
    \newline\newline
    \textbf{Alt Text:} A multi-panel grid of plots showing how metrics such as complexity, coefficient MAE, R², MSE, and training time change with varying network size, maximum edge value, penalty parameter L, and training length. SINDyG generally achieves lower complexity and error across conditions.}
	\label{fig:sensitivity}
\end{figure*}

In the general case, the network size, connectivity, and dynamics are variable and randomly chosen. We used the data from general case to conduct sensitivity analysis and assess the robustness of SINDy and SINDyG methods. Figure~\ref{fig:sensitivity} presents the results for various performance metrics, including complexity, coefficient MAE, training time, training $r^2$ and MSE, as well as testing $r^2$ and MSE scores. 

The first column of figure~\ref{fig:sensitivity} examines the impact of increasing the number of oscillators in the network. As the network size grows, so does the number of connections and the complexity of the dynamical equations for each node. This leads to an expected increase in complexity for both methods. However, SINDyG consistently yields models with lower complexity, regardless of the network size. Additionally, the coefficient MAE decreases as the number of nodes increases, which is in line with the characteristics of coefficient MAE for larger sets of potential coefficients. Importantly, SINDyG consistently outperforms SINDy in terms of coefficient MAE, indicating closer alignment with the true coefficients. While the training time remains relatively consistent for both methods, the train R-squared and MSE worsen for SINDy as the number of nodes increases, whereas SINDyG maintains performance. This trend extends to the test MSE and R-squared, highlighting SINDyG's ability to generalize to new, unseen data, especially in larger networks. These results demonstrate that as network size and coupling complexity increase, SINDy’s sparse regression performance degrades leading to higher model complexity and error, whereas SINDyG maintains robustness under the same conditions.

The second column explores the effect of increasing the maximum allowed value for edge weights, resulting in larger coupling term coefficients in the discovered equations. SINDy's complexity increases dramatically with higher edge values, while SINDyG exhibits only a marginal increase. This discrepancy also leads to a substantial rise in coefficient MAE for the SINDy predicted model, indicating a significant deviation from the true coefficients. Furthermore, the training time for both methods noticeably increases as the coefficient values rise. In the third column, we vary the value of the hyperparameter ($L$) in the penalty matrix ($\mathbf{P}[j, k]$). $L$ affects only SINDyG, so the standard SINDy curves serve as an L-independent baseline. As $L$ increases, the results for SINDyG become more robust, leading to enhanced predictive power and lower model complexity. The fourth column focuses on the impact of varying the training length, effectively reducing the amount of training data available.  We observe that SINDy's performance deteriorates as the data size gets smaller. In both predicted models, when we increase the train data the results are more robust as the standard error decreases. 

Overall, SINDy struggles with larger graphs and high coupling values, and its predictive power diminishes with limited training data. In contrast, SINDyG demonstrates greater resilience and consistently superior performance across various scenarios. Furthermore, SINDyG demonstrates more robust performance given significantly smaller standard errors compared to SINDy. This is also evident by the summary results provided in Table~\ref{tab:performance_metrics}.

\begin{table}[h] 
	\centering
	\caption{Performance Metrics (Mean $\pm$ SE) of SINDy and SINDyG on 100 Graphs with 5 Oscillators for Different Graph Types}
	\label{tab:performance_metrics}
    \resizebox{\textwidth}{!}{
	\begin{tabular}{l|cc|cc} 
		\toprule
		\textbf{Metric} & \multicolumn{2}{c|}{\textbf{ER}} & \multicolumn{2}{c}{\textbf{SF}} \\
		& \textbf{SINDy} & \textbf{SINDyG} & \textbf{SINDy} & \textbf{SINDyG} \\
		\midrule
		Complexity & 76.470 $\pm$ 3.4615 & {\bf61.870 $\pm$ 0.8531} & 71.790 $\pm$ 2.9351 & {\bf59.410 $\pm$ 0.7261} \\
		Coefficient MAE & 0.0323 $\pm$ 0.0026 & {\bf0.0107 $\pm$ 0.0004} & 0.0267 $\pm$ 0.0015 & {\bf0.0086 $\pm$ 0.0003} \\
		Train\_time (s) & 0.1224 $\pm$ 0.0059 & {\bf 0.1244 $\pm$ 0.0049} & 0.1408 $\pm$ 0.0111 & {\bf 0.1393 $\pm$ 0.0096} \\
		Train\_r2 & 0.9937 $\pm$ 0.0029 & {\bf 0.9993 $\pm$ 0.0005} & 0.9884 $\pm$ 0.0041 & {\bf 0.9998 $\pm$ 0.0001} \\
		Train\_MSE & 0.0003 $\pm$ 0.0000 & {\bf 0.0000 $\pm$ 0.0000} & 0.0002 $\pm$ 0.0000 & {\bf 0.0000 $\pm$ 0.0000} \\
		Test\_r2 & 0.5861 $\pm$ 0.1143 & {\bf 0.9732 $\pm$ 0.0126} & 0.5973 $\pm$ 0.1257 & {\bf 0.9715 $\pm$ 0.0127} \\
		Test\_MSE & 0.0076 $\pm$ 0.0017 & {\bf 0.0003 $\pm$ 0.0001} & 0.0067 $\pm$ 0.0013 & {\bf 0.0013 $\pm$ 0.0012} \\
		\bottomrule
	\end{tabular}
    }
\end{table}

\subsubsection{Computational Complexity Analysis}

In this section, we analyze the computational complexity of our proposed method (SINDyG) for discovering governing equations in a networked setting. The key component of our method is the ridge-regression with the objective function shown in Eq~\ref{eq:sindyGOpt}. The first component of this 
 objective function is the library matrix $\mathbf{\Theta}(\mathbf{X})$, which is calculated using $C$ candidate functions at $T$ time samples. This step typically costs  $\mathcal{O}(TC)$, though details may vary depending on the complexity of each candidate function. To incorporate network structure, we compute the penalty factor $\mathbf{P}[c,k]$ for each state variable $k$ and library term $c$. This requires calculating mean connectivity $m_{c k}$, which has the order $\mathcal{O}(KC)$, where $K$ represents the number of state variables. Once these two terms are calculated the objective function will be reduced to a standard ridge regression problem by scaling $\mathbf{\Theta}(\mathbf{X})$ and $\Xi$. These diagonal transformations require $\mathcal{O}(TC)$ and $\mathcal{O}(C)$, so they do not affect the leading term of the overall complexity. The dominant computational expense arises from solving the ridge-regression system, which entails forming the normal equations 
$(\mathbf{\Theta}^\top \mathbf{\Theta} + \lambda I)$ and factoring a 
$C\times C$ matrix. This procedure is well-known to cost $\mathcal{O}(TC^2 + C^3)$ flops~\cite{golub2013matrix}. 
Since STLSQ repeats thresholding and then solves $\tau$ times, the computation cost for $K$ state variables is formally:

\begin{equation}
    \mathcal{O}(K\sum_{l=1}^{\tau}{({TC_l}^{2}+{C_l}^3})),
\end{equation}

where $C_1 \geq C_2 \geq \cdots \geq C_\tau$. 

In the worst case, if thresholding removes almost nothing, we have $C_\tau \approx C$, giving the worst case computational cost of: 
\begin{equation}
    \mathcal{O}(\tau K({TC}^{2}+{C}^3)).
\end{equation}

However, in practice, $C_l$ can drop substantially after just a few iterations. This can lead to actual run-time being significantly less than the worst-case bound.

\section{Conclusion}

The main contribution of this study is the development of a novel method called SINDyG, which integrates network structure into sparse regression for the identification of governing equations in graph-structured dynamical systems. While SINDy has been applied to general and networked systems, SINDyG extends this framework by integrating known network topology directly into the sparse regression step. This graph-informed penalty improves model discovery by guiding the regression toward structurally plausible interactions. As demonstrated in our experiments, this approach leads to lower complexity, better predictive accuracy, and improved robustness across varying network sizes and coupling strengths. Recent work in symbolic regression like PySR~\cite{cranmer} and benchmark studies such as SRBench~\cite{SRBench} have demonstrated strong performance in discovering interpretable equations across a variety of systems. Integrating our graph-structured sparsity regularization into such solvers could improve robustness, especially in noisy or real-world settings. We view SINDyG as a first step toward a broader class of network-informed discovery algorithms, and we expect future work to build on this foundation.

The primary limitation of the proposed method lies in the selection and handling of appropriate basis or library functions, which could be challenging for complex systems with stochastic behavior. The size of the library functions in each point of the modeling has impact on the computational complexity of the model, which could be managed using regularization parameters. To address this, advanced optimization techniques, including sparsity-promoting algorithms and parallel computing frameworks, can be employed to reduce the computational burden without sacrificing accuracy. Additionally, exploring domain-specific knowledge to choose from library functions for specific applications, such as neuroscience, can improve computational efficiency. Lastly, while the method’s performance was validated using the Stuart–Landau model, further research is needed to assess the effectiveness of the proposed approach on other nonlinear dynamical systems, such as the Kuramoto and FitzHugh–Nagumo models. Nonetheless, the proposed graph-informed penalty term is very general and could be integrated with other symbolic regression algorithms to better understand and predict the behavior of complex systems across various domains, including neuroscience, ecology, finance, and epidemiology.



\section{Code and Data availability}
All the code and data used in this study are available at: 
\href{https://github.com/3sigmalab/SINDyG}{https://github.com/3sigmalab/SINDyG}

\bibliographystyle{comnet}

\bibliography{sample}

\begin{thebibliography}{00}

\bibitem{I97}
Becker, J., Brackbill, D. {\&} Centola, D. (2017)  Network dynamics of social influence in the wisdom of crowds. {\em Proceedings of the national academy of sciences}, \textbf{114}(26), E5070--E5076.

\bibitem{I3}
Bongard, J. {\&} Lipson, H. (2007)  Automated reverse engineering of nonlinear dynamical systems. {\em Proceedings of the National Academy of Sciences}, \textbf{104}(24), 9943--9948.

\bibitem{paper1}
Brunton, S.~L., Proctor, J.~L. {\&} Kutz, J.~N. (2016)  Discovering governing equations from data by sparse identification of nonlinear dynamical systems. {\em Proceedings of the national academy of sciences}, \textbf{113}(15), 3932--3937.

\bibitem{sindypic}
Champion, K., Brunton, S.~L. {\&} Kutz, J.~N. (2018)  Discovery of Nonlinear Multiscale Systems: Sampling Strategies and Embeddings. {\em arXiv preprint arXiv:1805.07411}.

\bibitem{cranmer}
Cranmer, M. (2023)  Interpretable machine learning for science with PySR and SymbolicRegression. jl. {\em arXiv preprint arXiv:2305.01582}.

\bibitem{SRBench}
de~Franca, F.~O., Virgolin, M., Kommenda, M., Majumder, M.~S., Cranmer, M., Espada, G., Ingelse, L., Fonseca, A., Landajuela, M., Petersen, B., Glatt, R., Mundhenk, N., Lee, C.~S., Hochhalter, J.~D., Randall, D.~L., Kamienny, P., Zhang, H., Dick, G., Simon, A., Burlacu, B., Kasak, J., Machado, M., Wilstrup, C. {\&} Cavaz, W. G.~L. (2024)  SRBench++: Principled Benchmarking of Symbolic Regression With Domain-Expert Interpretation. {\em IEEE Transactions on Evolutionary Computation}, pages 1--1.

\bibitem{Pysindy}
de~Silva, B.~M., Champion, K., Quade, M., Loiseau, J.-C., Kutz, J.~N. {\&} Brunton, S.~L. (2020)  Pysindy: a python package for the sparse identification of nonlinear dynamics from data. {\em arXiv preprint arXiv:2004.08424}.

\bibitem{delabays2024hypergraph}
Delabays, R., De~Pasquale, G., D{\"o}rfler, F. {\&} Zhang, Y. (2024)  Hypergraph reconstruction from dynamics. {\em arXiv e-prints}, pages arXiv--2402.

\bibitem{SL1}
Doelling, K.~B. {\&} Assaneo, M.~F. (2021)  Neural oscillations are a start toward understanding brain activity rather than the end. {\em PLoS biology}, \textbf{19}(5), e3001234.

\bibitem{ER}
Erdos, P., R{\'e}nyi, A.  et~al. (1960)  On the evolution of random graphs. {\em Publ. math. inst. hung. acad. sci}, \textbf{5}(1), 17--60.

\bibitem{SF}
Goh, K.-I., Kahng, B. {\&} Kim, D. (2001)  Universal behavior of load distribution in scale-free networks. {\em Physical review letters}, \textbf{87}(27), 278701.

\bibitem{golub2013matrix}
Golub, G.~H. {\&} Van~Loan, C.~F. (2013)  Matrix computations, 4th. {\em Johns Hopkins}.

\bibitem{Lasso}
Hastie, T., Tibshirani, R. {\&} Wainwright, M. (2015) {\em Statistical learning with sparsity: the lasso and generalizations}.
CRC press.

\bibitem{Kaheman}
Kaheman, K., Kutz, J.~N. {\&} Brunton, S.~L. (2020)  SINDy-PI: a robust algorithm for parallel implicit sparse identification of nonlinear dynamics. {\em Proceedings of the Royal Society A: Mathematical, Physical and Engineering Sciences}, \textbf{476}(2242), 20200279.

\bibitem{I7}
Kevrekidis, I.~G., Gear, C.~W., Hyman, J.~M., Kevrekidis, P.~G., Runborg, O., Theodoropoulos, C.  et~al. (2003)  Equation-free, coarse-grained multiscale computation: enabling microscopic simulators to perform system-level analysis. {\em Commun. Math. Sci}, \textbf{1}(4), 715--762.

\bibitem{I5}
Koza, J. (1992)  On the programming of computers by means of natural selection. {\em Genetic programming}.

\bibitem{brain}
Li, R., Sun, C., Dong, M., Wang, M., Gao, Q. {\&} Liu, X. (2024)  The Controllability Analysis of Brain Networks During Rhythmic Propagation. {\em IEEE Transactions on Network Science and Engineering}, \textbf{11}(4), 3812--3823.

\bibitem{I44}
Loiseau, J.-C. {\&} Brunton, S.~L. (2018)  Constrained sparse Galerkin regression. {\em Journal of Fluid Mechanics}, \textbf{838}, 42--67.

\bibitem{I2}
Marx, V. (2013)  The big challenges of big data. {\em Nature}, \textbf{498}(7453), 255--260.

\bibitem{powergrid}
Nakarmi, U., Rahnamay-Naeini, M. {\&} Khamfroush, H. (2019)  Critical component analysis in cascading failures for power grids using community structures in interaction graphs. {\em IEEE Transactions on Network Science and Engineering}, \textbf{7}(3), 1079--1093.

\bibitem{oscilatory}
Napoli, N.~J., Demas, M., Stephens, C.~L., Kennedy, K.~D., Harrivel, A.~R., Barnes, L.~E. {\&} Pope, A.~T. (2020)  Activation complexity: A cognitive impairment tool for characterizing neuro-isolation. {\em Scientific Reports}, \textbf{10}(1), 3909.

\bibitem{nijholt2022emergent}
Nijholt, E., Ocampo-Espindola, J.~L., Eroglu, D., Kiss, I.~Z. {\&} Pereira, T. (2022)  Emergent hypernetworks in weakly coupled oscillators. {\em Nature communications}, \textbf{13}(1), 4849.

\bibitem{SL}
Qin, Y., Menara, T., Bassett, D.~S. {\&} Pasqualetti, F. (2021)  Phase-amplitude coupling in neuronal oscillator networks. {\em Physical Review Research}, \textbf{3}(2), 023218.

\bibitem{I4}
Schmidt, M. {\&} Lipson, H. (2009)  Distilling free-form natural laws from experimental data. {\em science}, \textbf{324}(5923), 81--85.

\bibitem{I8}
Sugihara, G., May, R., Ye, H., Hsieh, C.-h., Deyle, E., Fogarty, M. {\&} Munch, S. (2012)  Detecting causality in complex ecosystems. {\em science}, \textbf{338}(6106), 496--500.

\bibitem{topal2023reconstructing}
Topal, I. {\&} Eroglu, D. (2023)  Reconstructing network dynamics of coupled discrete chaotic units from data. {\em Physical Review Letters}, \textbf{130}(11), 117401.

\bibitem{zhang2019convergence}
Zhang, L. {\&} Schaeffer, H. (2019)  On the convergence of the SINDy algorithm. {\em Multiscale Modeling \& Simulation}, \textbf{17}(3), 948--972.

\bibitem{traffic}
Zhang, Q., Yu, K., Guo, Z., Garg, S., Rodrigues, J. J. P.~C., Hassan, M.~M. {\&} Guizani, M. (2022)  Graph Neural Network-Driven Traffic Forecasting for the Connected Internet of Vehicles. {\em IEEE Transactions on Network Science and Engineering}, \textbf{9}(5), 3015--3027.

\bibitem{SR3}
Zheng, P., Askham, T., Brunton, S.~L., Kutz, J.~N. {\&} Aravkin, A.~Y. (2018)  A unified framework for sparse relaxed regularized regression: SR3. {\em IEEE Access}, \textbf{7}, 1404--1423.

\end{thebibliography}

\end{document}